\documentclass[twocolumn,twocolappendix]{aastex631}
\usepackage[encapsulated]{CJK}

\usepackage{amsmath}
\usepackage{xspace}

\newcommand{\HST}{HST\xspace}
\newcommand{\JWST}{JWST\xspace}

\newcommand{\versgrizli}{1.8.16.dev12\xspace}

\newcommand{\um}{\ensuremath{\mu\mathrm{m}}\xspace}
\newcommand{\unit}[1]{\ensuremath{\,\mathrm{#1}}\xspace}

\newcommand{\lmstar}{\ensuremath{\log_{10}(M_*/M_{\odot})}\xspace}

\newcommand{\grizli}{\texttt{grizli}\xspace}
\newcommand{\astrodrizzle}{astrodrizzle\xspace}
\newcommand{\prospector}{\texttt{Prospector}\xspace}

\newcommand{\galfit}{GALFIT\xspace}

\newcommand{\re}{\ensuremath{R_E}\xspace}

\newcommand{\reS}{\ensuremath{R_{E,S}}\xspace}
\newcommand{\nS}{\ensuremath{n_{S}}\xspace}

\newcommand{\nSLW}{\ensuremath{n_{S, \mathrm{F444W}}}\xspace}

\newcommand{\Conc}{\ensuremath{C}\xspace}

\newcommand{\recorr}{\ensuremath{R_{E}}\xspace}
\newcommand{\recorrLW}{\ensuremath{R_{E,\mathrm{LW}}}\xspace}
\newcommand{\recorrSW}{\ensuremath{R_{E,\mathrm{SW}}}\xspace}

\newcommand{\ConcLW}{\ensuremath{C_{\mathrm{F444W}}}\xspace}

\newcommand{\recorrfour}{\ensuremath{R_{E,\mathrm{F444W}}}\xspace}
\newcommand{\recorrfourkpc}{\ensuremath{R_{E,\mathrm{F444W,kpc}}}\xspace}

\newcommand{\logsizeratio}{\ensuremath{\log_{10}(\recorrLW/\recorrSW)}\xspace}

\newcommand{\submmfluxdens}{\ensuremath{S_{\rm1.2mm,int}}\xspace}

\begin{document}

\title{UNCOVER: 
The rest ultraviolet to near infrared multiwavelength structures and dust distributions of sub-millimeter-detected galaxies in Abell 2744}

\author[0000-0002-0108-4176]{Sedona H. Price}
\affiliation{Department of Physics and Astronomy and PITT PACC, University of Pittsburgh, Pittsburgh, PA 15260, USA}

\author[0000-0002-1714-1905]{Katherine A. Suess}
\affiliation{Department for Astrophysical \& Planetary Science, University of Colorado, Boulder, CO 80309, USA}

\author[0000-0003-2919-7495]{Christina C. Williams}
\affiliation{NSF?s National Optical-Infrared Astronomy Research Laboratory, 950 N. Cherry Avenue, Tucson, AZ 85719, USA}
\affiliation{Steward Observatory, University of Arizona, 933 North Cherry Avenue, Tucson, AZ 85721, USA}

\author[0000-0001-5063-8254]{Rachel Bezanson}
\affiliation{Department of Physics and Astronomy and PITT PACC, University of Pittsburgh, Pittsburgh, PA 15260, USA}

\author[0000-0002-3475-7648]{Gourav Khullar}
\affiliation{Department of Physics and Astronomy and PITT PACC, University of Pittsburgh, Pittsburgh, PA 15260, USA}

\author[0000-0002-7524-374X]{Erica J. Nelson}
\affiliation{Department for Astrophysical and Planetary Science, University of Colorado, Boulder, CO 80309, USA}

\author[0000-0001-9269-5046]{Bingjie Wang (\begin{CJK*}{UTF8}{gbsn}???\ignorespacesafterend\end{CJK*})}
\affiliation{Department of Astronomy \& Astrophysics, The Pennsylvania State University, University Park, PA 16802, USA}
\affiliation{Institute for Computational \& Data Sciences, The Pennsylvania State University, University Park, PA 16802, USA}
\affiliation{Institute for Gravitation and the Cosmos, The Pennsylvania State University, University Park, PA 16802, USA}

\author[0000-0003-1614-196X]{John R. Weaver}
\affiliation{Department of Astronomy, University of Massachusetts, Amherst, MA 01003, USA}

\author[0000-0001-7201-5066]{Seiji Fujimoto}
\altaffiliation{NHFP Hubble Fellow}
\affiliation{Department of Astronomy, The University of Texas at Austin, Austin, TX 78712, USA}

\author[0000-0002-5588-9156]{Vasily Kokorev}
\affiliation{Kapteyn Astronomical Institute, University of Groningen, 9700 AV Groningen, The Netherlands}

\author[0000-0002-5612-3427]{Jenny E. Greene}
\affiliation{Department of Astrophysical Sciences, Princeton University, 4 Ivy Lane, Princeton, NJ 08544}

\author[0000-0003-2680-005X]{Gabriel Brammer}
\affiliation{Cosmic Dawn Center (DAWN), Niels Bohr Institute, University of Copenhagen, Jagtvej 128, K{\o}benhavn N, DK-2200, Denmark}

\author[0000-0002-7031-2865]{Sam E. Cutler}
\affiliation{Department of Astronomy, University of Massachusetts, Amherst, MA 01003, USA}

\author[0000-0001-8460-1564]{Pratika Dayal}
\affiliation{Kapteyn Astronomical Institute, University of Groningen, 9700 AV Groningen, The Netherlands}

\author[0000-0001-6278-032X]{Lukas J. Furtak}\affiliation{Physics Department, Ben-Gurion University of the Negev, P.O. Box 653, Be?er-Sheva 84105, Israel}

\author[0000-0002-2057-5376]{Ivo Labbe}
\affiliation{Centre for Astrophysics and Supercomputing, Swinburne University of Technology, Melbourne, VIC 3122, Australia}

\author[0000-0001-6755-1315]{Joel Leja}
\affiliation{Department of Astronomy \& Astrophysics, The Pennsylvania State University, University Park, PA 16802, USA}
\affiliation{Institute for Computational \& Data Sciences, The Pennsylvania State University, University Park, PA 16802, USA}
\affiliation{Institute for Gravitation and the Cosmos, The Pennsylvania State University, University Park, PA 16802, USA}

\author[0000-0001-8367-6265]{Tim B. Miller}
\affiliation{Center for Interdisciplinary Exploration and Research in Astrophysics (CIERA) and Department of Physics \& Astronomy, Northwestern University, IL 60201, USA}

\author[0000-0003-2804-0648 ]{Themiya Nanayakkara}
\affiliation{Centre for Astrophysics and Supercomputing, Swinburne University of Technology, PO Box 218, Hawthorn, VIC 3122, Australia}

\author[0000-0002-9651-5716]{Richard Pan}\affiliation{Department of Physics and Astronomy, Tufts University, 574 Boston Ave., Medford, MA 02155, USA}

\author[0000-0001-7160-3632]{Katherine E. Whitaker}
\affiliation{Department of Astronomy, University of Massachusetts, Amherst, MA 01003, USA}
\affiliation{Cosmic Dawn Center (DAWN), Denmark}

\begin{abstract}
With the wavelength coverage, sensitivity, and high spatial resolution of \JWST, it is now possible to peer through the dust attenuation to probe the rest-frame near infrared (NIR) and stellar structures of extremely dusty galaxies at cosmic noon ($z\sim1-3$). In this paper we leverage the combined ALMA and JWST/HST coverage in Abell 2744 to study the multiwavelength ($0.5-4.4\um$) structures of 11 sub-millimeter (sub-mm) detected galaxies at $z\sim0.9-3.5$ that are fainter than bright ``classical'' sub-mm galaxies (SMGs); 7 of which are detected in deep X-ray data. While these objects reveal a diversity of structures and sizes, all are smaller and more concentrated towards longer wavelengths. Of the X-ray-detected objects, only two show evidence for appreciable AGN flux contributions (at $\gtrsim2\um$). 
Excluding the two AGN-dominated objects, 
the smaller long wavelength sizes indicate that their rest-frame NIR light profiles, inferred to trace their stellar mass profiles, are more compact than their optical profiles. The sub-mm detections and visible dust lanes suggest centrally-concentrated dust is a key driver of the observed color gradients. 
Further, we find that more concentrated galaxies tend to have lower
size ratios (rest-frame NIR to optical); 
this suggests that the galaxies with the most compact light distributions also have the most concentrated dust. The 1.2mm flux densities and size ratios of these 9 objects suggest that both total dust quantity and geometry impact these galaxies' multiwavelength structures. Upcoming higher resolution 1.2mm ALMA imaging will facilitate joint spatially-resolved analysis and will directly test the dust distributions within this representative sub-mm population.
\end{abstract}


\section{Introduction} \label{sec:intro}

The universe looks very different in the sub-millimeter (sub-mm) than at shorter wavelengths. Galaxies which are bright in the sub-mm are often faint or undetected at optical or near infrared (NIR) wavelengths, as their high dust content attenuates the shorter wavelength light and then re-emits this energy as thermal continuum in sub-mm to millimeter (mm) wavelengths (e.g., \citealt{2014PhR...541...45C}). 
These sub-mm detected galaxies are typically distant ($z\gtrsim1$), massive, dusty, and star-forming (e.g., \citealt{2014PhR...541...45C}, \citealt{2017MNRAS.471.2453S}, \citealt{2020MNRAS.494.3828D}; \citealt{2020RSOS....700556H}). 
However, their detailed rest-frame  NIR structures, 
which contain key information about their formation, 
remained uncertain with previously available observations.

The sensitivity of sub-mm and mm interferometers, particularly ALMA, has opened a window to directly study the impacts of dust well below the limits of ultraluminous infrared galaxies (ULIRGs) and classical bright sub-millimeter galaxies (SMGs), revealing dust continuum and obscured star formation even in ``main sequence'' star-forming galaxies. 
Probing the faint sub-mm population requires deep field or lensing cluster surveys, and there have been relatively limited ALMA studies in the $\sim0.1-1\unit{mJy}$ flux density regime to date (e.g., \citealt{2016ApJS..222....1F}, \citealt{2016ApJ...833...68A}, \citealt{2020ApJ...897...91G}, \citealt{2022A&A...658A..43G}, \citealt{MunozArancibia23}). 
However, these surveys have shown that the dust content in the fainter population is still significant, with $\sim85\%$ of star formation at cosmic noon ($z\sim1-3$) obscured by dust (e.g., \citealt{2017MNRAS.466..861D}). 
Thus, a full picture of galaxy assembly at cosmic noon requires an understanding of these lower sub-mm luminosity sources.

While the detailed substructures of bright dusty objects such as SMGs have been accessible with ALMA and NOEMA (revealing merging starbursts as well as disks; e.g., \citealt{2016ApJ...833..103H, 2019ApJ...876..130H}, \citealt{2016ApJ...833...12R}, \citealt{2018ApJ...863...56C}, \citealt{2019ApJ...879...54L}, \citealt{2019ApJ...877L..23P}, \citealt{2020ApJ...901...74T}), reaching the fainter population at high spatial resolution to map the dust continuum has remained a challenge, requiring very long exposures even with ALMA's sensitivity. Little is known about this population's intrinsic structures, including their distribution of dust and active star formation, leaving us without a complete picture of how this faint sub-mm population builds up or whether this population represents a particular evolutionary phase in galaxy formation.

An alternative, complementary avenue to characterize the dust distribution and obscured structures in faint sub-mm sources is using high-resolution rest-optical imaging to probe the stellar distribution directly. 
Prior to JWST, the high spatial resolution and sensitivity \emph{Hubble Space Telescope} (HST) enabled detailed structural measurements, but its limited wavelength coverage (up to $\sim1.6\um$) probed only as far as the rest frame optical at cosmic noon. 
And while \emph{Spitzer} had longer wavelength coverage to probe the rest-frame near infrared (NIR) at these redshifts, it
lacked the spatial resolution necessary to map stellar structures. 
Consequently, past structural measurements have typically been based on one HST filter at the rest-optical. 
However, dust complicates the interpretation of galaxies' rest-optical structures. 
In particular, dust (as well as age and metallicity, which also contribute to radial color gradients) has been shown to result in dramatically different stellar size estimates depending on the dust concentration and distribution, as probed by relative differences between half-mass and half-light sizes (e.g., \citealt{2012ApJ...753..114W}, \citealt{2016ApJ...822L..25L, 2017ApJ...844L...2L}, \citealt{2019ApJ...877..103S}, 
\citealt{Miller23}, 
\citealt{2023MNRAS.524.4128Z}). 
Mock observations have also found dust geometry to be a major complication in structural interpretation of dusty galaxies (\citealt{2022MNRAS.510.3321P},  \citealt{2023MNRAS.524.4128Z}).

With the launch of JWST \citep{2023PASP..135f8001G}, it is now possible to obtain high resolution imaging over an expanded wavelength range from $\lesssim1\um$ out to $\sim4.4\um$ (corresponding to the rest-frame $\sim0.3-1.4\um$ at $z\sim1-3$), providing information across the full stellar spectral energy distribution (SED). 
Critically, this coverage includes the rest-frame NIR probing the largely unattenuated stellar continuum.
In addition to enabling robust measurements of sizes and stellar densities, comparing the rest-frame NIR with the rest-frame UV/optical (``color gradients'') allows us to constrain radial variations in mass-to-light ratios. These color gradients can be driven by non-uniform dust distributions or variations in relative stellar ages and/or metallicities; therefore, color gradients also indicate where active growth occurs in galaxies now relative to their past (e.g., \citealt{Suess21, Suess22, Suess23}). 
Early results from JWST point to dust distribution as a major contributor to observed color profiles (e.g., \citealt{Miller22}; also \citealt{Suess22}) and without the simultaneous rest-NIR coverage, intrinsic size estimates may be biased towards larger values (e.g., \citealt{2023MNRAS.524.4128Z}, among others). 

By combining JWST/NIRCam imaging 
with existing shorter wavelength HST imaging, we can now perform detailed multiwavelength structural studies of the distribution of stars, dust, and star formation in extremely dusty distant galaxies, revealing new insights about 
the nature and formation of this population. 
A number of studies have already leveraged early JWST observations to constrain the structures of distant dusty galaxies, including both sub-mm detected galaxies (including \citealt{2022ApJ...936L..19C, 2023ApJ...942L..19C}, \citealt{Gillman23}, \citealt{Kokorev23}, \citealt{2023ApJ...948L...8R},  
\citealt{Smail23}, \citealt{Wu23}, 
\citealt{2023arXiv230816895B}) and other dusty galaxies (e.g., 
\citealt{Gomez-Guijarro23}, 
\citealt{2023arXiv230707599L}, 
\citealt{2023arXiv230510944L}, 
\citealt{Nelson23},  \citealt{2023arXiv230519331M}). 
Building upon these works, we study the structures of a sample of galaxies within a cluster field, where the strong lensing boost allows us to examine the relation between total dust content and detailed multiwavelength structures for a variety of galaxies pushing to faint intrinsic fluxes and compact physical sizes.

Here we present a multiwavelength ($0.5-4.4\um$) structural analysis of 11 $z\sim0.9-3.5$ sub-mm detected 
($S_{\mathrm{1.2mm,int}}\sim 0.2-1.5\unit{mJy}$) 
galaxies in the Abell 2744 cluster field as measured with \HST and \JWST. 
This sample includes all galaxies detected in $1.2\unit{mm}$ ALMA continuum observations of the Abell 2744 primary cluster core from the ALMA Lensing Cluster Survey (ALCS; \citealt{Fujimoto23}) 
and ALMA-Hubble Frontier field (ALMA-HFF; \citealt{MunozArancibia23}) programs. 
We visually examine the rest-optical and NIR structures, then quantify the multiwavelength structures by fitting single 
component 
models and deriving residual-corrected flux profiles 
in every filter for the full sample. 
We also calculate the ratio of sizes between long ($>4.4\um$) and short ($\sim1.5-2.7\um$) wavelengths, 
which is sensitive to color gradient strengths probing radial changes in dust, age, or metallicity.  
Dust is expected to be an important driver of these color gradients, 
given this sample is sub-mm detected and a subset exhibits signatures consistent with dust lanes or patchy dust distributions. 
Finally, we examine the structural measurements together with the intrinsic, lensing-corrected 1.2mm flux densities, to probe the impact of both dust quantity and geometry on these galaxies' multiwavelength structures.

Throughout this work we adopt a flat $\Lambda$CDM cosmology with $\mathrm{H_0=70\,km\,s^{-1}\,Mpc^{-1}}$, $\Omega_M=0.3$, and $\Omega_{\Lambda}=0.7$.

\begin{figure*}[ht!]
\centering
\includegraphics[width=\textwidth]{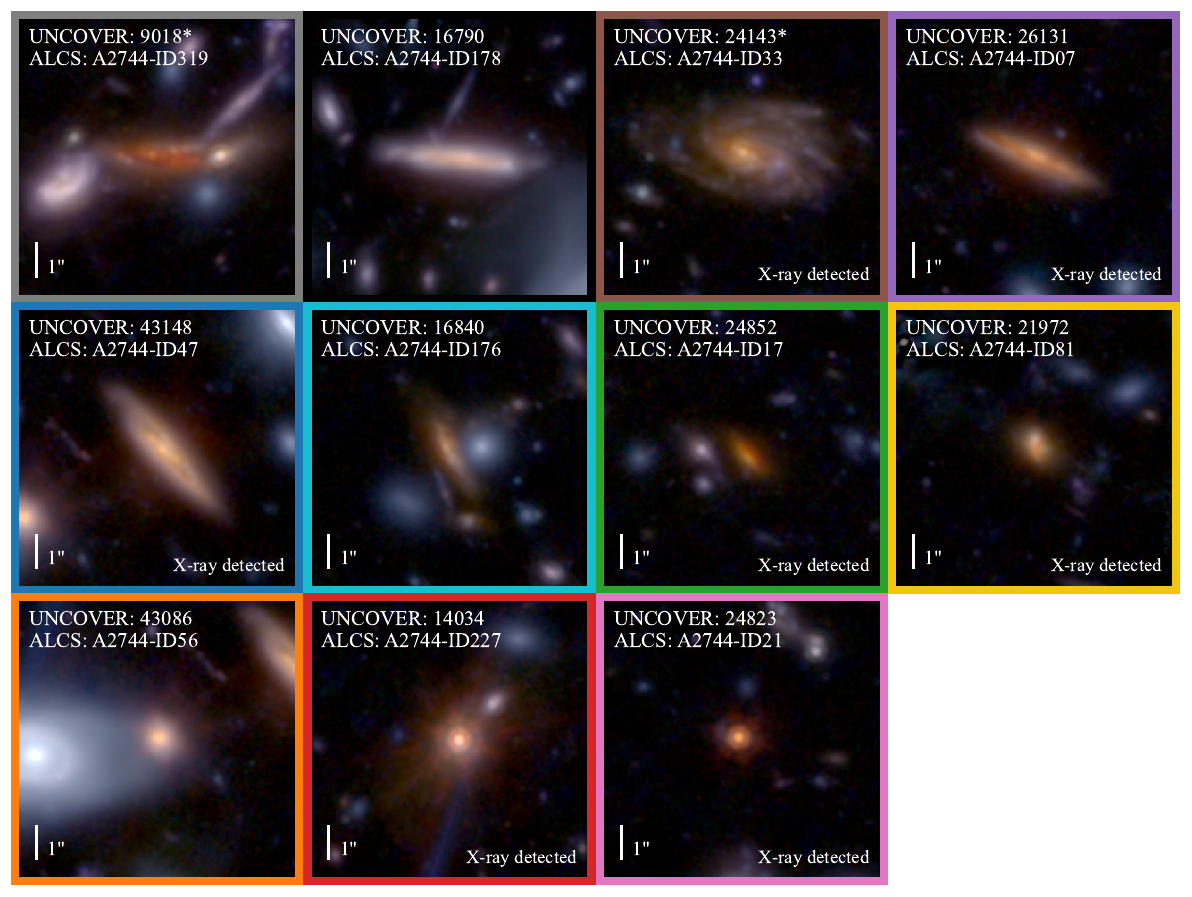}
\caption{RGB composite images (F115W+F150W/F200W+F277W/F356W+F410M+F444W; PSF-matched to F444W) of the sample, in decreasing order of F444W residual-corrected effective radii (\recorr). 
The galaxy IDs from UNCOVER (\citealt{Weaver23}) and 
ALCS (\citealt{Fujimoto23}) are 
noted in each panel, and the outlines show the colors used to denote each galaxy throughout the paper.  
Asterisks on the UNCOVER IDs 
(i.e., 9018, 24143) 
indicate that the alternative, less aggressively-deblended segmentation was used for structural modeling (see Sec.~\ref{sec:data}). 
Objects with X-ray detections are noted.
} 
\vspace{2mm}
\label{fig:sampleRGB}
\end{figure*}

\vspace{20pt}

\section{Data}
\label{sec:data}

To study the rest-frame UV to near infrared (NIR) structures 
of sub-mm detected galaxies at cosmic noon ($z\sim1-3$), 
we leverage the combination of low spatial resolution ALMA continuum observations (beam FWHM $\sim0.9\arcsec\times0.7\arcsec$; \citealt{Fujimoto23}) with deep, high spatial resolution \JWST/\HST coverage available for the Abell 2744 cluster field. 
We select galaxies from ALCS (\citealt{Kokorev22}, \citealt{Sun22}, \citealt{Fujimoto23}; including ALMA-HFF observations; \citealt{2017A&A...597A..41G}, \citealt{MunozArancibia23}), as presented in \citet{Fujimoto23}. 
We include both sources in the ALCS 1.2mm continuum blind ($\mathrm{SNR_{natural\,map}}\geq5$; $N=6$) and IRAC-prior ($4.0\leq \mathrm{SNR_{natural\,map}}< 5$ and $\mathrm{SNR_{IRAC,CH2}}\geq 5$; $N=5$) detection catalogs, for a total of 11 galaxies.

For this analysis, we use all 
public \JWST/NIRCam (F115W, F150W, F200W, F277W, F356W, F410M, F444W) 
and \HST/ACS (F435W, F606W, F814W) and WFC3 (F105W, F125W, F140W, F160W) imaging covering Abell 2744. 
Mosaics for all filters are produced with a common WCS grid using \grizli (\versgrizli; \citealt{Brammer19}, \citealt{Kokorev22}) and \astrodrizzle \citep{Gonzaga12}, with 
JWST/NIRCam short wavelength (SW; F115W, F150W, F200W) filters drizzled onto a 20mas scale, while JWST/NIRCam long wavelength (LW; F277W, F356W, F410M, F444W) and HST filters are drizzled onto a 40 mas scale.\footnote{\url{https://s3.amazonaws.com/grizli-v2/JwstMosaics/v7/index.html}} 
See \citet{Bezanson22} for full details of the imaging reduction and mosaicing. 
Background subtraction and modeling and subtraction of bright cluster galaxies is then performed, 
as detailed in \citet{Weaver23}.

\begin{figure*}[t!]
\vspace{5mm}
\centering
\includegraphics[width=\textwidth]{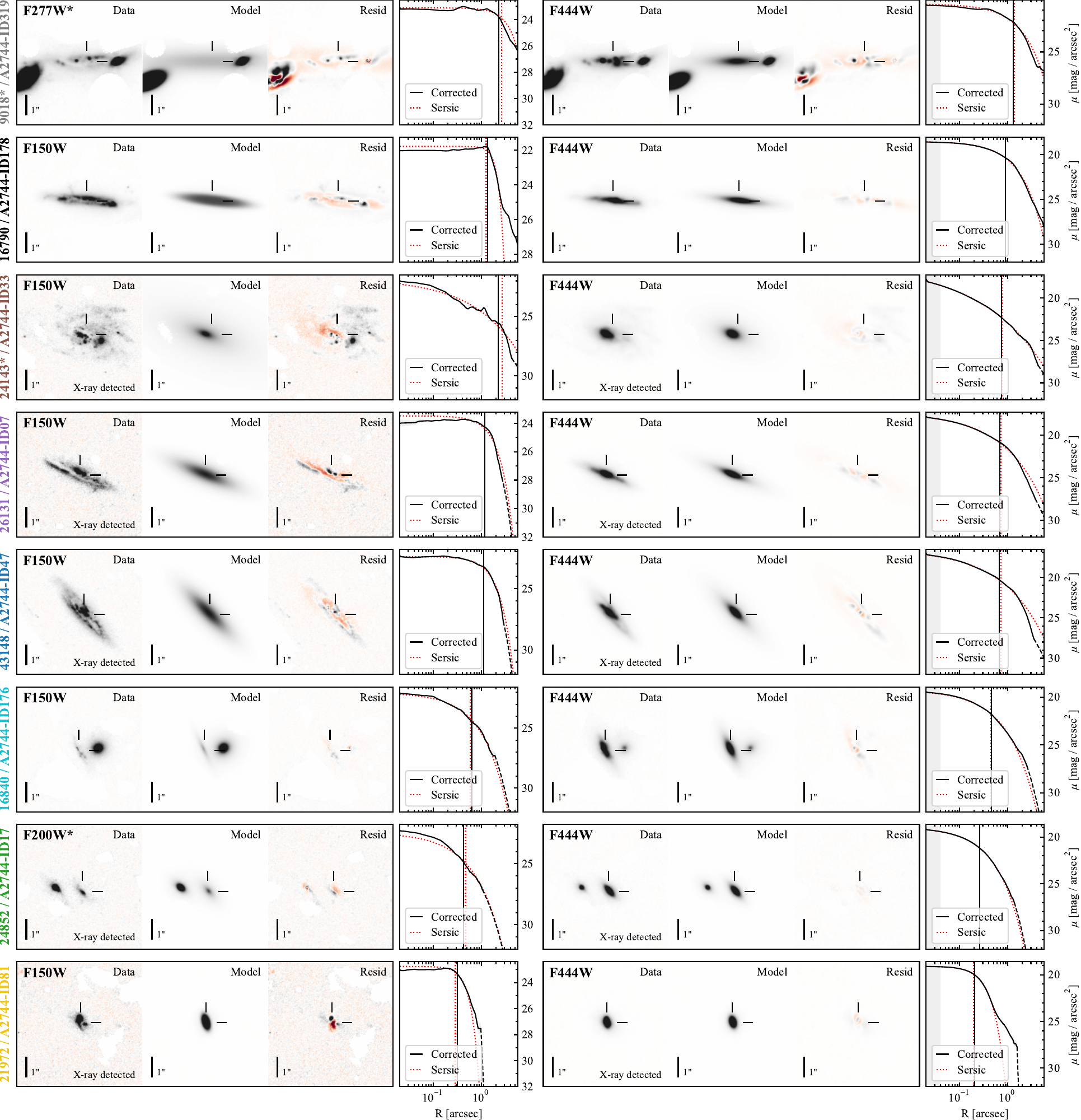}
\caption{
Example structural modeling for a short (\emph{left}; F150W if fit and unflagged, otherwise F200W/F277W) and long wavelength band (\emph{right}; F444W) for each galaxy. 
For each object and filter, we show postage stamps of the galaxy, the best-fit \galfit model, and the residual (\emph{leftmost 3 panels}, respectively), using the same linear color scale for all three images 
and cropped to 75\% of the fit extent (i.e., $6{\arcsec}\times6{\arcsec}$) for clarity. 
The residual-corrected and best-fit S\'ersic model surface brightness (SB) profiles are also shown (black and red dashed lines, respectively; \emph{rightmost panel}), with the effective radii of each curve (\recorr, \reS) denoted by the corresponding vertical line. For the residual-corrected SB profile, the uncertainty is shown with the shaded black region, and the dashed portion of the curve marks where the extrapolated S\'ersic profile is adopted due to limited signal-to-noise. 
The vertical gray region indicates radii smaller than the 
pixel scale. Objects with X-ray detections are noted in the lower right of the data image panels.
} 
\vspace{5mm}
\label{fig:SW_LW_fits}
\end{figure*}

\renewcommand{\thefigure}{2}

\begin{figure*}[t!]
\centering
\includegraphics[width=\textwidth]{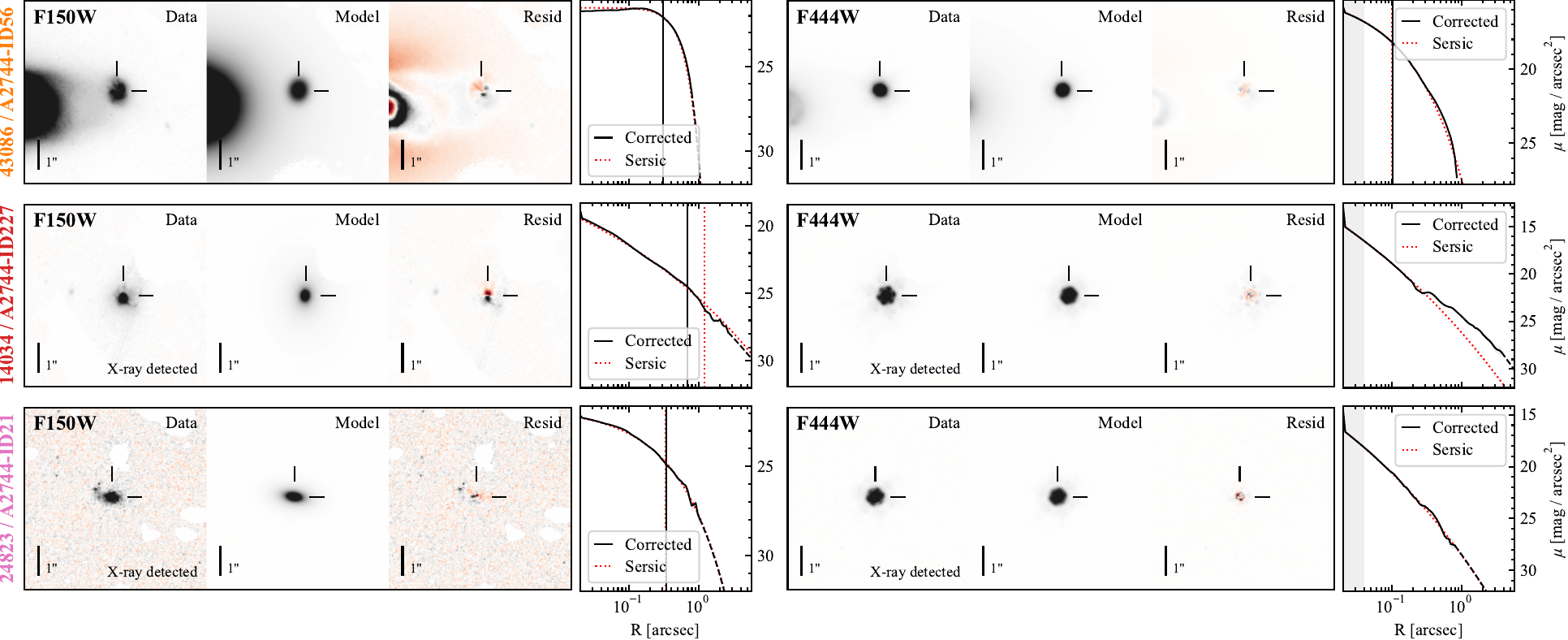}
\caption{(cont.)} 
\label{fig:SW_LW_fits_cont}
\vspace{2mm}
\end{figure*}

\renewcommand{\thefigure}{\arabic{figure}}
\setcounter{figure}{2}

The ALCS sources are matched to \JWST-detected sources from  the UNCOVER photometric catalog (DR2; \citealt{Weaver23}) with a search radius of $0.5\arcsec$. 
NIR counterparts for all ALCS sources are identified. 
However, in two cases (UNCOVER IDs 
9018, 24143; 
denoted throughout with an asterisk), visual inspection reveals the galaxies are shredded into multiple sources in the UNCOVER catalog as a result of the detection deblending thresholds, which were optimized to detect faint, high-redshift sources. 
As the photometry and segmentation maps are used for 
simultaneous modeling or masking of neighboring objects in the structural modeling (see Sec.~\ref{sec:methods}), 
for these two objects we instead perform 
an alternative detection (using SEP; \citealt{Barbary18}) with less aggressive deblending parameters 
($\texttt{KERNEL=3.5}$ pixel FWHM Gaussian, $\texttt{MINAREA=3}$ pixels, 
$\texttt{THRESH=0.8}\,\sigma$, 
$\texttt{DEBLEND\_NTHRESH=8}$, 
$\texttt{DEBLEND\_CONT=0.0001}$, 
$\texttt{CLEAN=Y}$, 
$\texttt{CLEAN\_PARAM=1.0}$), 
and use the resulting segmentation maps and photometric parameters (derived using the same photometric pipeline as \citealt{Weaver23} except for the parameter changes) 
for structural modeling of these two galaxies. 
We calculate the magnifications ($\mu$) for each object from the updated lensing model by \citet[v1.1]{Furtak23}, adopting the best-available redshifts. 
We also estimate stellar masses from the HST and JWST photometry, as well as the 1.2mm ALMA fluxes (from the most recent DUALZ catalog; \citealt{Fujimoto23b}) using the same procedure as the UNCOVER DR2 stellar population synthesis (SPS) catalog \citep{Wang23}, with the following modifications. 
First, we fix the redshifts to $z_{\rm spec}$ or $z_{\rm gris}$ when available. 
Second, we use the alternative, less deblended photometry measured for the aforementioned two objects. For the single object with only a $z_{\rm phot}$ (16840), we use the redshift and stellar mass directly from the DR2 SPS catalog \citep{Wang23}. 
For clarity, we emphasize that all physical properties (i.e., stellar mass and physical size) and the intrinsic 1.2mm flux densities are corrected for gravitational lensing using the magnification $\mu$.

We use deep Chandra X-ray maps of Abell 2744 (2.1Ms, presented in \citealt{Bogdan24}, considering both blind and IRAC-prior detections) to determine if any of these objects host AGN. 
We find that the majority of our sample is X-ray detected (7/11: 24143*, 26131, 43148, 24852, 21972, 14034, 24823), likely hosting 
AGN. 
This suggests that the high AGN fraction amongst somewhat brighter sub-mm selected galaxies \citep[e.g.,][]{Serjeant10,Shim22} 
continues even towards lower sub-mm fluxes. 
This X-ray detection fraction is even higher than found by previous studies for brighter sub-mm-detected galaxies \citep[e.g.,][]{Laird10,Johnson13, Wang13}.\footnote{We defer any comparison of AGN and X-ray-detection fractions among faint ($\lesssim1\unit{mJy}$) sub-mm-detected galaxies compared to brighter sub-mm-detected populations to future analysis.} 
Objects which are X-ray detected are noted in figures throughout this work. 
As discussed later (see Sec.~\ref{sec:methods}, Appendix~\ref{sec:appendixA}), 
we determine that the host galaxy dominates the light even out to $4.4\um$ in 5 of these objects (24143*, 26131, 43148, 24852, 21972); 
we therefore infer that these objects host highly obscured AGN, such that we can still analyze the wavelength-dependent host galaxy morphology. 
We infer that the other 2 objects (14034, 24823) 
host dust-reddened AGN, which start to appreciably contribute to the total flux at $\gtrsim2\um$.

Color composite \JWST images for our sample are shown in Fig.~\ref{fig:sampleRGB}, 
and the sample IDs, positions, best-available 
redshifts, magnifications, and stellar masses 
are listed in Table~\ref{table:tab1}.
For structural modeling, we use empirical PSFs constructed for each band from unsaturated stars within the mosaic, 
renormalized so the energy enclosed within $4\arcsec$ aligns with typical calibration levels (see \citealt{Weaver23} for full details).

\vfill \eject 

\section{Methods}
\label{sec:methods}

To measure the multiwavelength structural parameters of our sample, 
we perform single S\'ersic component fits in the image plane 
using \galfit \citep{Peng02,Peng10} 
on $8\arcsec\times8\arcsec$ cutouts 
(from the original resolution, non-PSF matched images)
in all filters where the integrated galaxy flux has $\mathrm{SNR}\geq10$. 
First, we fit the parameters in F444W, 
using initial values based on the detection catalog 
and adopting parameter limits on the 
S\'ersic index ($\nS=[0.2,8]$), 
major axis effective radius ($\reS=[0.3,300]\unit{pixels}$), and 
axis ratio ($b/a=[0.05,1]$), 
while requiring the total magnitude to be within $\pm3\unit{mag}$ 
and the center position to within $\pm10$ pixels of the initial values. 
We simultaneously fit any neighboring objects with $\mathrm{min}(\Delta\mathrm{mag_X})<1.5$ (considered over all filters X) and with any part of their segmentation maps falling within $1.5\arcsec$ of the primary object, 
and mask all other detected objects are using the detection segmentation maps. 
We then fit in all other filters in which the galaxy has $\mathrm{SNR}\geq10$, 
fixing the position and PA of all objects to the F444W values. 
For all filters, we flag the fit if \galfit returns a flag $>0$ or if one of the parameters reaches the enforced limit (with the exception that we do not flag fits where $\nS=8$ or $b/a=1$). 
Additionally, as some of our objects are nearly point sources in the reddest filters, we do not flag fits where only \reS has numerical stability issues and $\reS<1\unit{pixel}$.

We modify this procedure if $(b/a)_{\mathrm{F444W}}>0.75$, as the  position angle (PA) may be poorly constrained. In this case, we fix the positions to the best-fit $(x_c,y_c)_{\mathrm{F444W}}$ and perform an intermediate fit on the shortest-wavelength JWST filter with $\mathrm{SNR}\geq10$ to determine the galaxy's PA. 
After this SW filter fit, we refit F444W  with the positions and PA of all objects fixed, and then fit all remaining filters as in the normal procedure. 
We show example short and long wavelength \galfit model results for each object (F150W if fit and unflagged, otherwise F200W/F277W; and F444W) in the first three panels of the left and right columns in Fig.~\ref{fig:SW_LW_fits}.

As single S\'ersic component fits do not capture 
galaxy substructures (which are clearly present in many of the filters for galaxies in our sample), 
we also compute residual-corrected flux profiles and corrected effective radii \recorr 
for all filters 
following \citet{Szomoru10,Szomoru12}. 
We compute residual flux profiles from the \galfit model residual images (masking all neighboring objects) using elliptical annular apertures with 
fixed PA and axis ratio based on each band's best-fit values. 
We then determine uncertainties using derived per-band empty aperture 
scaling relations (to capture correlated noise; as in \citealt{Skelton14}). 
We show these residual-corrected and \galfit S\'ersic fit surface brightness profiles and respective effective radii (\recorr, \reS) for example short and long wavelength bands (F150W or F200W and F444W) of each object in the fourth panel of each column in Fig.~\ref{fig:SW_LW_fits}. 
We note that the parametric S\'ersic \reS are in fairly good agreement with the corrected \recorr 
(median \recorr/\reS = 1.01) 
suggesting that to at least first order the single-component fits capture the average light distributions fairly well, even in the presence of the observed substructures.

Because structural fitting is performed in the image plane, we must account for the lensing magnification on sizes. 
As our sample has only modest magnifications of $\mu\sim2-4$, we correct for lensing with a factor of $1/\sqrt{\mu}$. Throughout this paper, all physical sizes have been corrected for both cosmology and this lensing factor. Angular sizes, as measured directly in the image plane, are uncorrected for lensing.

As an alternative to the parametric S\'ersic index, we also derive the empirical concentration \Conc as defined in \citet{Conselice03} for all filters. 
We calculate the concentration from the residual-corrected flux profiles, in order to 
mitigate the impact of PSF-broadening (compared to measuring \Conc from the observed images).

We also investigate the impact of dust-reddened or dust-obscured AGN on the morphologies of the X-ray detected objects in our sample (7/11). 
We perform a second, alternative set of morphological fits, using the same methodology as above but adding a point source component. 
The point source-to-extended flux ratios from these fits are presented in Appendix~\ref{sec:appendixA}, 
and also discussed in Sec.~\ref{sec:results}. 
For the analysis that follows, we exclude objects with appreciable ($\gtrsim10\%$) estimated point-source AGN flux contribution, and use morphological parameters derived from the single-component S\'ersic fits in the analysis.

\begin{figure}[t!]
\vspace{-2mm}
\centering
\includegraphics[width=0.485\textwidth]{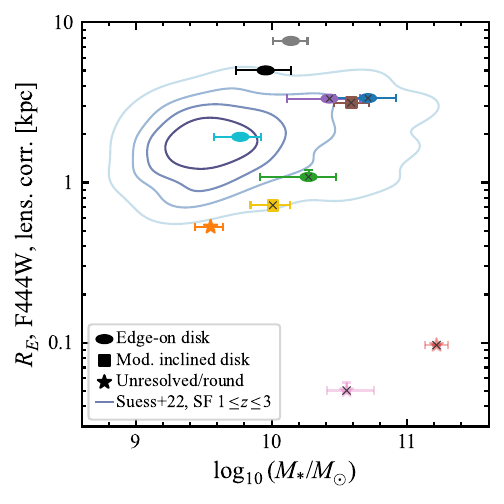}
\vglue -4mm
\caption{
Stellar mass versus semi-major effective radius from single-S\'ersic fits in F444W (corrected for lensing using a factor of $1/\sqrt{\mu}$). 
The symbols shapes denote the primary visual morphological categories of 
edge-on disks (ovals), moderately inclined disks (squares), and unresolved or small round sources (stars), 
with the same symbol color for each galaxy as in Figs.~\ref{fig:sampleRGB},~\ref{fig:structures_meas}. 
X-ray-detected objects are marked with gray Xs. 
Compared to the sample of star-forming galaxies at $z\sim1-3$ from \citet{Suess22}, 
roughly half of our sample lies within the scatter of mass versus size for ``normal'' star-forming galaxies. Three objects are notable outliers to larger (9018*) and smaller 
(14034, 24823; dust-reddened but AGN-dominated at long wavelengths; denoted with partial transparency) 
radii for their mass.
}
\label{fig:mstar_vs_re}
\end{figure}

\begin{figure*}[t!]
\vspace{-2mm}
\centering
\includegraphics[width=1.0\textwidth]{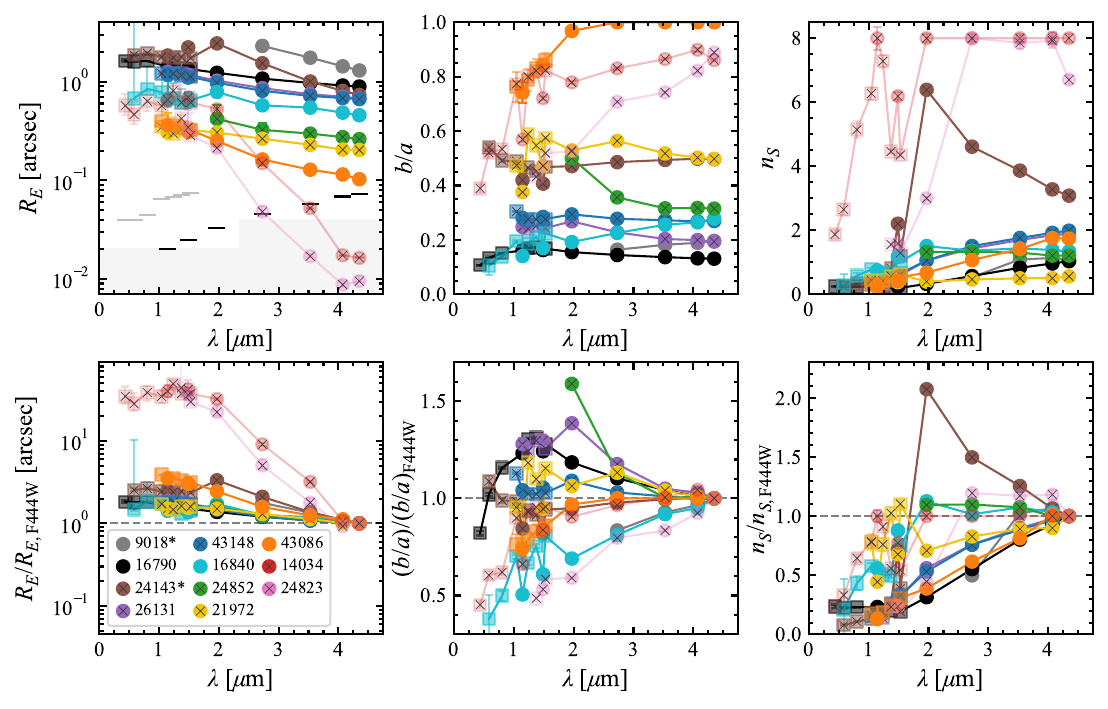}
\vglue -3mm
\caption{
Measured structural parameters for all bands with good (unflagged) fits. 
The top row shows the corrected effective radius (\recorr), 
axis ratio ($b/a$), 
and S\'ersic index (\nS) versus filter wavelength ($\lambda$) (\emph{left to right}, respectively). In the bottom row we plot the ratio of these values to the value measured from F444W. 
\JWST and \HST filters are denoted with solid circles and translucent squares, respectively, and in the upper left panel the horizontal black (gray) lines denote the HWHM of the \JWST (\HST) filter PSFs, and the filled light gray regions the pixel scale. X-ray-detected objects are marked with gray Xs. 
The galaxies show a trend of decreasing \recorr and increasing \nS towards longer wavelengths, consistent with a high central concentration of dust. Overall the axis ratios are broadly similar at all wavelengths, with the exception of three objects (including the two objects that are AGN-dominated at long wavelengths, denoted with partial transparency and shown here for reference) with the smallest \recorrLW that are very round ($b/a\gtrsim0.8$) at the longest wavelengths. 
} 
\vspace{2mm}
\label{fig:structures_meas}
\end{figure*}

\begin{figure*}[t!]
\centering
\vspace{-1mm}
\hglue -5pt
\includegraphics[width=1.02\textwidth]{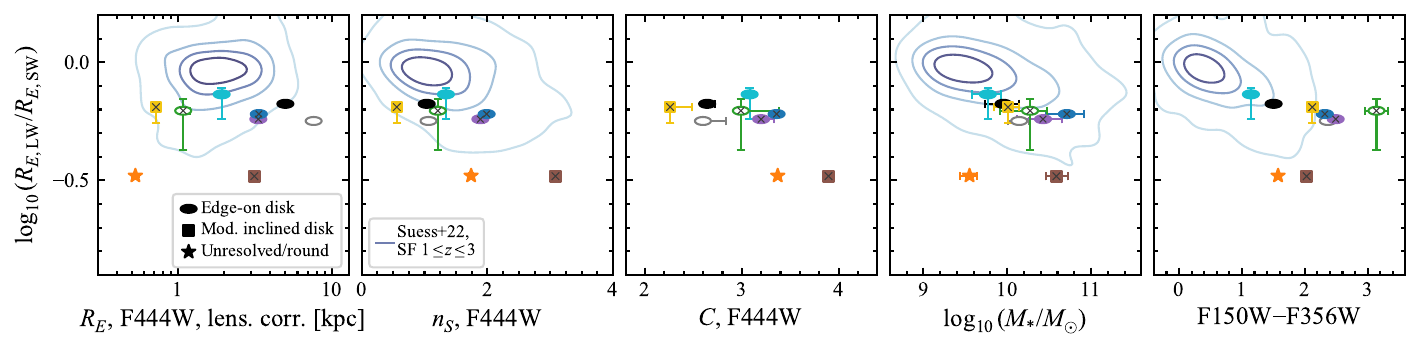}
\vglue -2mm
\caption{
Ratio of long (F444W) to short wavelength (F150W if sufficient SNR and unflagged, solid symbols; otherwise F200W 
or F277W, 
open symbols) 
sizes, 
\logsizeratio, 
as a function of structural and integrated color parameters. 
The symbols are the same as in Fig.~\ref{fig:mstar_vs_re} (including marking X-ray-detected objects with gray Xs).
We plot these ratios versus \recorr (lensing corrected), 
\nS, and the concentration $C$ (as measured from the corrected flux profiles; see Sec.~\ref{sec:methods}) from F444W (\emph{first}, \emph{second}, and \emph{third panels}, respectively). 
Additionally, we plot against the inferred stellar mass (\emph{fourth panel}), and against the observed integrated F150W--F356W color (\emph{fifth panel}), which approximates the rest-frame V--J color given the sample redshift ($z\sim2$). 
As a comparison, we also show the measured radii ratios \logsizeratio  
versus $R_{E,\mathrm{F444W}}$, \nSLW, \lmstar, 
and V--J 
from 
\citet[colored contours]{Suess22}. 
These comparisons demonstrate that our sample has size ratios 
(reflecting the color gradients and 
thus inferred dust attenuation gradients) extending from the range of ``typical'' $z\sim1-3$ galaxies down to steeper values for the more compact objects.
} 
\vspace{1mm}
\label{fig:structures_slopes}
\end{figure*}

\section{Results}
\label{sec:results}

Our multiwavelength $0.5-4.4\um$ imaging of these 11 ALMA-detected galaxies reveals a diversity of structures. 
Visual inspection (i.e., Figs.~\ref{fig:sampleRGB} \& \ref{fig:SW_LW_fits}) reveals galaxies that appear to be large disks, both edge-on (9018*, 16790, 26131, 43148; 
see also \citealt{Kokorev23} for other analysis of 9018*/ID319, which \citeauthor{Kokorev23} report may be in the early phases of a merger with the compact companion to the West given its undisturbed disk morphology) and more face-on (24143*; see also \citealt{Wu23}). 
The sample also includes other smaller galaxies that appear disk-like and edge-on (16840, 24852), 
a small galaxy that appears to be a moderately inclined disk with a misaligned dust lane but may be irregular (21972), 
and small objects which are irregular at short wavelengths but appear compact and round at long wavelengths 
(43086; also 14034, 24823; see Sec.~\ref{sec:methods}, Appendix~\ref{sec:appendixA}, and below).

Quantitatively, 
the sample also exhibits a wide range structural parameter values. 
In Fig.~\ref{fig:mstar_vs_re} we show \recorrfourkpc versus inferred stellar masses for our sample.
Our sample ranges from low ($\sim10^{9.5}{\rm{M_{\odot}}}$) to high ($\sim10^{11}{\rm{M_{\odot}}}$) stellar mass. 
We find some of the objects visually that are identified as large disks lie above the mass-size distribution of ``normal'' star-forming galaxies at $z\sim1-3$ from \citet{Suess22}, while the remainder of edge-on and moderately-inclined disks fall within the scatter of the \citeauthor{Suess22} sample. 
Unresolved/round objects include one galaxy (with the lowest stellar mass) just barely below the population scatter.

The two most extremely small objects (14034, 24823) host dust-reddened X-ray detected AGN.  Based on  alternative S\'ersic+point source morphological fits, these objects have appreciable ($\gtrsim10\%$) AGN flux contributions at long wavelengths ($\gtrsim2\um$; see Sec.~\ref{sec:methods} and Appendix~\ref{sec:appendixA}), and are therefore excluded from the later structural analysis. 
The other 5 X-ray-detected objects appear to have fully obscured AGN even at $4.4\um$ (i.e., minimal point source flux contribution), with the HST and JWST-derived morphology are dominated by the host galaxy (see Sec.~\ref{sec:methods} and Appendix~\ref{sec:appendixA}). 
These 5 objects are retained in the sample, and the morphological parameters derived from the single-S\'ersic fits are adopted for the remainder of this analysis.

In Fig.~\ref{fig:structures_meas}, we consider measured structural parameters (effective radius, axis ratio, and S\'ersic index) as a function of wavelength.  
We find sizes \recorr spanning $1-2$ orders of magnitude at a given wavelength, axis ratios $b/a$ spanning from $\sim0.1-1$, and S\'ersic indices ranging from 
$\nS\sim0.2-1$ in HST/ACS F435W to 
$\nS\sim0.5-8$ in JWST/NIRCam F444W. 
The three galaxies with the smallest \re in F444W ($\lesssim0.1\arcsec$) have sufficiently concentrated long-wavelength light distributions (as evidenced by their relatively high \nS) to exhibit visible diffraction spikes in the log-scale color images (Fig.~\ref{fig:sampleRGB}). 
We do note that the presented structural parameters are sensitive to a number of additional systematic uncertainties. 
Previous work has shown systematic uncertainties can be important for GALFIT-derived morphological parameters (e.g., \citealt{vanderWel12}), likely leading to underestimates of the total uncertainties presented in this work. 
We attempt to account for some of these systematic uncertainties (due to non-smooth galaxy light profiles) by using residual corrections to derive \recorr, though this does not provide corrections for other, directly-measured structural parameters (i.e., $b/a$, \nS).

We also observe common trends in the variation of the structural parameters with wavelength, as seen in Fig.~\ref{fig:structures_meas} (showing both the measured structural parameters, \emph{top}, and the trends normalized to the values in F444W, \emph{bottom}).
With increasing wavelength, we find 
\re decreases and \nS generally increases, with each galaxy appearing more compact and concentrated at longer wavelengths. 
(We find a similar trend of increasing \Conc with wavelength, in agreement with the \nS trend.) 
We also find the axis ratios do not drastically change with wavelength (i.e., varying only by $\sim0.1-0.2$), with the exception of three objects that are extremely compact at long wavelengths 
(43086; and the two hereafter excluded dust-reddened AGN, 14034, 24823). 

Changes in morphology with wavelength, or equivalently color gradients, reflect non-uniform dust distributions as well as relative age or metallicity changes. 
Color gradients thus provide a powerful probe of how galaxies assemble their stellar mass profiles (e.g., \citealt{Suess21,Suess23}). 
In these sub-mm detected galaxies, we infer that dust is a primary driver of the observed color gradients and morphological changes with wavelength, as we will discuss later. 
Thus to estimate the radial distribution of dust, we calculate a size ratio \logsizeratio  between a long ($>4\um$) and and short wavelength ($\sim1.5-2.7\um$) band, probing the rest-frame NIR and optical, respectively, for this sample. 
We use F444W as the LW band, and preferentially use F150W (if the galaxy has sufficient SNR and an unflagged fit in that band), or otherwise F200W or F277W, as the SW band. 
These size ratios are shown as a function of F444W \recorr in kpc 
(corrected for lensing using $1/\sqrt{\mu}$ as in Fig.~\ref{fig:mstar_vs_re}), 
\nS, and \Conc in Fig.~\ref{fig:structures_slopes}. 
Our sample has size ratios that range from relatively mild values ($\sim-0.2$), similar to those observed in $z\sim1-3$ \HST-selected star-forming galaxies by \citet{Suess22}, down to steeper ratios ($\sim-0.5$) for the more compact objects in F444W.

While we observe a correlation between \logsizeratio 
and \recorrfour, where the smallest objects have the most extreme gradients, we find more clear anti-correlations between \nSLW and 
\ConcLW and the size ratio. 
The \citet{Suess22} sample also show slight tails 
of stronger gradients towards lower \recorrfour  
and higher \nSLW, 
though the observed trend for the sub-mm detected galaxies extends far beyond the region spanned by ``typical'' SFGs.

We also consider the size ratios versus 
stellar mass, \lmstar (\emph{fourth panel}). 
We find that most (7/9) of our sample (excluding the two X-ray detected objects hosting dust-reddened AGN that contribute to the long-wavelength flux) roughly overlaps with the anti-correlation trend found by \citet{Suess22}, but the 
remaining two galaxies  
have lower size ratios at a fixed stellar mass. This suggests that stellar mass alone does not set galaxies' size ratios and that measures of long wavelength central concentration (i.e., \ConcLW or \nSLW) 
correlate more strongly with size ratio 
(in contrast to \citealt{Gomez-Guijarro23}, who find 
dust concentration correlates most strongly with mass).

Finally, we examine the size ratios as a function of 
the integrated observed F150W--F356W color, which 
approximates the rest-frame V--J color for our sample redshift ($z\sim2$; Fig.~\ref{fig:structures_slopes}, \emph{fifth panel}). 
Again, this is consistent with the anti-correlation of \logsizeratio with rest-frame V--J found by \citet{Suess22}. Our sample extends that trend to redder objects. 
In Fig.~\ref{fig:structures_slopes} we show that most (7/9) 
of our sample continues the \logsizeratio{} -- V--J trend shown by the \citeauthor{Suess22} contours; however, our galaxy sample also includes two objects that have lower size ratios than expected given their color (as also seen with stellar mass).

Rest-frame UVJ colors 
have been proposed to help distinguish between color gradients that are driven by dust versus age \citep[e.g.,][]{Miller22, Miller23}, though \citealt{Leja19} caution that interpretation of spatially-resolved UVJ colors with respect to dust gradients requires further testing. This analysis relies on the fact that 
more dust tends to produce redder U--V colors \emph{and} redder V--J colors, where as older ages tend to produce redder U--V colors but bluer V--J colors. To test our rest-frame color gradients, 
we use EAzY-Py (\citealt{Brammer2008}; using \texttt{tweak\_fsps\_QSF\_12\_v3} templates) to interpolate, or for 4 objects, slightly extrapolate, 
the residual-corrected deconvolved profiles 
(as in Fig.~\ref{fig:SW_LW_fits}) 
to construct rest-frame UVJ color profiles. 
We find that all galaxies in our sample have both redder U--V and V--J colors in the centers 
than the outskirts, consistent with increasing dust attenuation towards their centers. 
Combined with the sub-mm detection, the visible signatures consistent with dust lanes or patchy dust distributions in some objects, and the emergence of a stellar bar at long wavelengths for at least one object (43148, which cannot be explained by stellar age gradients), these signatures are consistent with high central dust concentrations being an important driver of the observed color gradients. 
Accurately detangling the contribution of dust attenuation and age to the observed color profiles will require pixel-to-pixel resolved SED fitting. 
Future high-resolution sub-mm observations (e.g., with a forthcoming ALMA Cycle 10 program; PI: V. Kokorev) and spatially-resolved spectroscopy 
will also provide valuable constraints on the dust attenuation and age profiles.

\begin{figure}[t!]
\centering
\hglue -5pt
\includegraphics[width=0.495\textwidth]{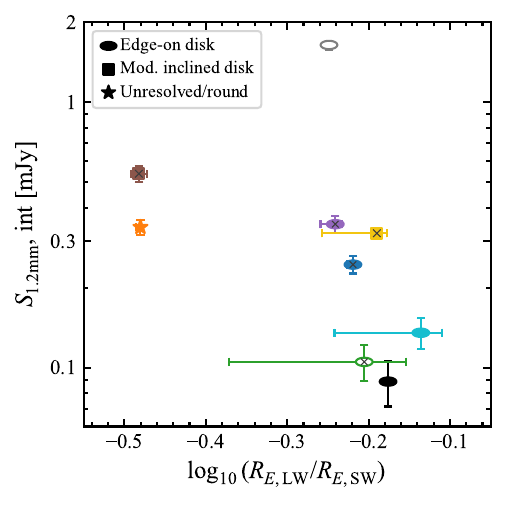}
\vglue -2mm
\caption{
Comparison of the intrinsic, lensing-corrected ALMA 1.2mm flux density 
to the 
ratio of sizes 
\logsizeratio (using the same symbol definitions as in Fig.~\ref{fig:structures_slopes}). 
We find that our sample exhibits a rough anti-correlation between  
sub-mm flux density and size ratio, as might be expected from varying the total dust quantity at fixed star-dust (or dust-star-AGN) geometry. 
However, the sample exhibits notable scatter, including a clear outlier (9018*); this suggests that both dust geometry and total quantity impact their size ratios.
The uncertain fraction of 
AGN-driven emission 
to the 1.2mm flux densities in the 5 X-ray detected, highly obscured AGN host galaxies complicates the interpretation of sub-mm fluxes 
with respect to galaxy structures. 
However, even after excluding these sources, the distribution suggests that the inference that both dust quantity and geometry impact size ratios holds. 
Future analysis with high-resolution sub-mm maps will disentangle contributions from AGN emission and non-dust drivers of color gradients.
} 
\label{fig:submmflux_vs_structures}
\end{figure}

\section{Discussion \& Conclusions}
\label{sec:disc_concl}

This sample demonstrates that the population of sub-mm detected galaxies is not homogeneous in structure. 
Among our 11 objects, we observe disk-like structures over a range of inclinations and sizes, as well as smaller, more irregular objects that also host extremely compact central components at the longest wavelengths. 
In the rest-frame NIR (see Fig.~\ref{fig:SW_LW_fits}), these structures are consistent with stellar disks (with a range of central concentrations; 8/11), small spheroidal or face-on stellar disks (1/11), and NIR continuum dominated by heavily 
dust-reddened AGN that have appreciable AGN flux contributions at long wavelengths (2/11; excluded from the full analysis).

Even with this morphological diversity, we find evidence that within individual galaxies the dust and NIR continuum light distributions are similar. 
Specifically, we find the size ratios --- and thus color gradients --- correlate with size and concentration in F444W.  
As we infer that the color gradients are primarily driven by 
dust, based on their sub-mm detection, visible dust lane and patchy distributions, and inferred rest-frame UVJ color profiles (Sec.~\ref{sec:results}), this correlation suggests that the objects with smaller and more concentrated NIR continuum light (assumed to trace the stars, for all but the 2 excluded dust-reddened X-ray-detected AGN) also have smaller and more compact dust distributions, while objects with less concentrated NIR light also have less concentrated dust. 
We see further evidence for self-similarity between the NIR continuum and dust distributions from the trend of axis ratios with wavelength: for most objects there is little change in $b/a$ with wavelength, suggesting similar dust geometry to the underlying stellar distribution. 
For the two excluded dust-reddened X-ray detected AGN, 
the axis ratio changes reflect the emergence of these 
point-source AGN as they become less obscured at longer wavelengths, 
also suggesting distribution self-similarity in order to obscure the central AGN.

Finally, the distribution of 
1.2mm flux densities (using updated fluxes from the recently-released DUALZ catalog by \citealt{Fujimoto23b}; Fig.~\ref{fig:submmflux_vs_structures}) versus size ratios suggests that both dust geometry and total dust quantity (as potentially evidenced by sub-mm flux) are important in setting these galaxies' multiwavelength structures. 
For fixed dust-star geometry, increasing the total dust will tend to result in stronger color gradients and more extreme short-to-long wavelength size ratios. Increasing the total dust for a fixed distribution (i.e., increasing the dust column densities) will preferentially attenuate the inner regions compared to the outskirts (assuming dust column density $\sim A_{\rm V}$; \citealt{Salim20}), increasing the effective radii at short wavelengths while minimally impacting the long-wavelength sizes (assuming the dust is not optically thick at those wavelengths). This would tend to produce an anti-correlation between the sub-mm flux densities and size ratios.

Variations in dust-star (or dust-star-AGN) geometry --- including patchiness in the dust distribution --- could change the relation between total dust quantity and color gradient, introducing scatter between the sub-mm flux densities and size ratios. Furthermore, if a galaxy is optically thick even in the rest-frame NIR, dust attenuation would also increase the long-wavelength effective radii relative to the stellar half-mass radii. This could result in larger LW-to-SW size ratios (as both the SW and LW effective radii are larger than the stellar half-mass radius), compared to objects where only the SW is attenuated while the LW is relatively unattenuated. 

Our sample exhibits both some degree of anti-correlation between \submmfluxdens and \logsizeratio and notable scatter (including a strong outlier, 9018*, which may be optically thick even in F444W, potentially explaining the unexpectedly high size ratio given the total sub-mm flux density as discussed above). Taken together, we infer that both total dust quantity and dust geometry are important in explaining size ratios. 
However, we note that non-dust drivers of color gradients (e.g., age gradients) could also introduce scatter. Future analysis of high-resolution maps of the dust continuum, and spatially-resolved stellar population synthesis modeling, will reveal the relative impact of dust quantity and geometry on multiwavelength morphologies and color gradients, and disentangle the impact of non-dust effects.

Furthermore, high (low) sub-mm fluxes can also reflect high (low) incident radiation from star formation or AGN activity. 
The uncertain AGN contributions to the 1.2mm flux density for the 5 X-ray detected but highly obscured AGN host galaxies retained in this analysis (24143*, 26131, 43148, 24852, 21972) thus complicates the above inferences regarding dust content and dust geometry. 
Nonetheless, the non-X-ray-detected objects (9018*, 16790, 16840, 43086) span much of the parameter space that might be expected for both total dust and dust geometry contributions to setting the observed size ratios --- suggesting that the conclusion that both dust quantity and geometry are important holds even with the high AGN fraction in this sample.  
High-resolution dust continuum maps will additionally help address this issue, by spatially detangling the central AGN contribution from the extended emission driven by star formation radiation.

This analysis highlights the power of using  high-resolution, deep NIR imaging from JWST to unravel the multi-wavelength, multi-component structures of extremely dusty galaxies, which were previously shrouded from view. 
More detailed observations and modeling to directly recover the stellar mass and dust distributions are necessary to better understand the structures of sub-mm detected galaxies. 
This type of spatially-resolved spectral energy distribution (SED) modeling would benefit dramatically from higher-resolution ALMA continuum observations, which are forthcoming in an approved ALMA Cycle 10 program (2023.1.00626.S; PI V. Kokorev). 
Resolved NIR spectroscopy (either from a slit or IFU) would also be critical to directly probe the reddening and thus 
disentangle the dust, age, and metallicity color degeneracies. 
Future work with these observations will also determine how well dust and stellar geometries are captured through  multiwavelength structural measurements similar to those presented in this work.
Finally, JWST and ALMA will continue to map larger fields, including the recently completed ALMA Cycle 9 program DUALZ covering the full extended Abell 2744 cluster \citep{Fujimoto23b}, yielding significantly larger samples that fully span the demographics of dust-obscured/sub-mm-detected galaxies across cosmic time. 
Larger samples would enable full explorations of sub populations and constrain their respective evolutionary pathways. Combined with parallel efforts to characterize structures of less extreme galaxy populations, these studies will continue bridging the gap between optical studies of relatively low dust galaxies and sub-mm studies of the most dusty galaxies.


\section*{}
\noindent
We thank the anonymous referee for a constructive report which has improved this manuscript. 
This work is based in part on observations made with the NASA/ESA/CSA \emph{James Webb Space Telescope}. 
These observations are associated with JWST-GO-2561, JWST-ERS-1324, and JWST-DD-2756. Support for program JWST-GO-2561 was provided by NASA through a grant from the Space Telescope Science Institute, which is operated by the Associations of Universities for Research in Astronomy, Incorporated, under NASA contract NAS5-26555. This research is also based on observations made with the NASA/ESA \emph{Hubble Space Telescope} obtained from the Space Telescope Science Institute, which is operated by the Association of Universities for Research in Astronomy, Inc., under NASA contract NAS 5?26555. These observations are associated with programs HST-GO-11689, HST-GO-13386, HST-GO/DD-13495, HST-GO-13389, HST-GO-15117, and HST-GO/DD-17231. 
All of the data presented in this paper were obtained from the Mikulski Archive for Space Telescopes (MAST) at the Space Telescope Science Institute. The specific observations used to produce these mosaics can be accessed via \dataset[10.17909/zn4s-0243]{http://dx.doi.org/10.17909/zn4s-0243}.
This paper makes use of the ALMA data: ADS/JAO. ALMA \#2018.1.00035.L, \#2019.1.00237.S, \#2021.1.01246.S, and \#2021.1.00407.S. ALMA is a partnership of the ESO (representing its member states), NSF (USA) and NINS (Japan), together with NRC (Canada), MOST and ASIAA (Taiwan), and KASI (Republic of Korea), in cooperation with the Republic of Chile. The Joint ALMA Observatory is operated by the ESO, AUI/NRAO, and NAOJ.
This research was supported in part by the University of Pittsburgh Center for Research Computing, RRID:SCR\_022735, through the resources provided. Specifically, this work used the H2P cluster, which is supported by NSF award number OAC-2117681.
VK acknowledges funding from the Dutch Research Council (NWO) through the award of the Vici Grant VI.C.212.036.
PD acknowledges support from the Dutch Research Council (NWO) through the award of the VIDI Grant 016.VIDI.189.162 (``ODIN") and the European Commission's and University of Groningen's CO-FUND Rosalind Franklin program. 

\vspace{5mm}
\facilities{\JWST(NIRCam), \HST(ACS and WFC3), ALMA}

\software{
    astropy \citep{astropy:2013, astropy:2018, astropy:2022},  
    EAzY-Py \citep{Brammer2008}, 
    \galfit \citep{Peng02, Peng10}, 
    matplotlib \citep{matplotlib:2007},  
    numpy \citep{numpy:2020}, 
    scipy \citep{scipy:2020}, 
    seaborn \citep{seaborn:2021}, 
    photutils \citep{photutils:1.8.0}, 
    \prospector \citep{Johnson2021}
}


\setlength{\tabcolsep}{3.5pt}

\begin{deluxetable*}{ccccccccccccc}
\tablecolumns{13}
\tablecaption{Sample properties and structural parameters
\label{table:tab1}}
\tablehead{
\colhead{ID}  
& \colhead{ID$_{\rm ALCS}$} 
& \colhead{RA}  
& \colhead{Dec}
& \colhead{$z_{\mathrm{best}}$} 
& \colhead{Type} 
& \colhead{Ref.} 
& \colhead{$\mu$}
& \colhead{$\log_{10}\!M_*$}
& \colhead{$R_{E,\mathrm{F444W}}$} 
& \colhead{$n_{S,\mathrm{F444W}}$} 
& \colhead{\ConcLW} 
& \colhead{$\log_{10}\!\left(\frac{\recorrLW}{\recorrSW}\right)$} 
\\
\colhead{(1)} 
& \colhead{(2)} 
& \multicolumn{2}{c}{(3)} 
& \colhead{(4)} 
& \colhead{(5)} 
& \colhead{(6)} 
& \colhead{(7)} 
& \colhead{(8)} 
& \colhead{(9)} 
& \colhead{(10)} 
& \colhead{(11)} 
& \colhead{(12)} 
}
\startdata 
\phm{*}9018* & ID319 & 3.575986 & -30.413174 & 2.580 & 1 & 3,9 & 1.85 & $10.1_{-0.1}^{+0.1}$ & $1.298_{-0.003}^{+0.004}$ & 1.06$\pm$0.01 & $2.598_{-0.002}^{+0.237}$ & $-0.249_{-0.001}^{+0.001}$,\emph{\scriptsize{a}}\\[1pt]
16790 & ID178 & 3.600396 & -30.396138 & 0.940 & 2 & 1,5 & 1.98 & $10.0_{-0.2}^{+0.2}$ & $0.892_{-0.001}^{+0.002}$ & 1.03$\pm$0.00 & $2.642_{-0.001}^{+0.077}$ & $-0.176_{-0.001}^{+0.001}$\phm{,\emph{\scriptsize{a}}}\\[1pt]
\phm{*}24143* & ID33 & 3.584924 & -30.381780 & 3.060 & 1 & 1,2,9 & 3.19 & $10.6_{-0.1}^{+0.1}$ & $0.733_{-0.013}^{+0.013}$ & 3.08$\pm$0.01 & $3.896_{-0.009}^{+0.010}$ & $-0.482_{-0.009}^{+0.010}$\phm{,\emph{\scriptsize{a}}}\\[1pt]
26131 & ID07 & 3.579685 & -30.378407 & 2.410 & 1 & 1,2,9 & 2.70 & $10.4_{-0.3}^{+0.2}$ & $0.675_{-0.007}^{+0.010}$ & 1.89$\pm$0.01 & $3.199_{-0.002}^{+0.137}$ & $-0.241_{-0.018}^{+0.008}$\phm{,\emph{\scriptsize{a}}}\\[1pt]
43148 & ID47 & 3.571959 & -30.382986 & 1.670 & 1 & 7 & 2.80 & $10.7_{-0.2}^{+0.2}$ & $0.664_{-0.004}^{+0.004}$ & 1.99$\pm$0.01 & $3.359_{-0.001}^{+0.001}$ & $-0.219_{-0.007}^{+0.004}$\phm{,\emph{\scriptsize{a}}}\\[1pt]
16840 & ID176 & 3.572349 & -30.395966 & $3.65_{-0.14}^{+0.09}$ & 3 & 6 & 2.92 & $9.8_{-0.2}^{+0.2}$ & $0.456_{-0.004}^{+0.020}$ & 1.34$\pm$0.01 & $3.081_{-0.010}^{+0.010}$ & $-0.136_{-0.106}^{+0.026}$\phm{,\emph{\scriptsize{a}}}\\[1pt]
24852 & ID17 & 3.581273 & -30.380227 & 3.475 & 1 & 7,8 & 3.14 & $10.3_{-0.4}^{+0.2}$ & $0.261_{-0.003}^{+0.026}$ & 1.21$\pm$0.00 & $2.990_{-0.010}^{+0.393}$ & $-0.205_{-0.166}^{+0.051}$,\emph{\scriptsize{a}}\\[1pt]
21972 & ID81 & 3.582499 & -30.385459 & 3.056 & 1 & 7,8 & 4.65 & $10.0_{-0.2}^{+0.1}$ & $0.203_{-0.001}^{+0.006}$ & 0.56$\pm$0.00 & $2.259_{-0.002}^{+0.230}$ & $-0.190_{-0.068}^{+0.012}$\phm{,\emph{\scriptsize{a}}}\\[1pt]
43086 & ID56 & 3.573258 & -30.383501 & 1.500 & 1 & 1,2,9 & 2.70 & $9.6_{-0.1}^{+0.1}$ & $0.103_{-0.000}^{+0.000}$ & 1.74$\pm$0.03 & $3.371_{-0.001}^{+0.001}$ & $-0.480_{-0.006}^{+0.001}$\phm{,\emph{\scriptsize{a}}}\\[1pt]
14034 & ID227 & 3.568930 & -30.402792 & 2.580 & 2 & 1,5 & 1.85 & $11.2_{-0.1}^{+0.1}$ & $0.016_{-0.000}^{+0.000}$ & 8.00$\pm$0.13 & $5.808_{-0.000}^{+0.000}$ & $-1.625_{-0.020}^{+0.020}$\phm{,\emph{\scriptsize{a}}}\\[1pt]
24823 & ID21 & 3.592096 & -30.380472 & 2.640 & 1 & 1,2,9 & 2.28 & $10.6_{-0.1}^{+0.2}$ & $0.010_{-0.000}^{+0.001}$ & 6.70$\pm$0.14 & $5.310_{-0.000}^{+0.000}$ & $-1.558_{-0.213}^{+0.072}$\phm{,\emph{\scriptsize{a}}}\\[1pt]
\enddata
\tablecomments{
(1) UNCOVER ID (DR2; \citealt{Weaver23}). 
(2) ALCS ID \citep{Fujimoto23}, omitting the prefix `A2744'.
(3) UNCOVER catalog coordinate (J2000).
(4) Best available redshift.
(5) Redshift type: 1: spectroscopic, 2: HST grism, 3: photometric. 
(6) References: 
1: \citet{Fujimoto23}, 
2: \citet{MunozArancibia23}, 
3: \citet{Kokorev23}, 
4: \citet{Laporte17}, 
5: \citet{Wang15},
6: \citet[][UNCOVER, DR2 SPS catalog]{Wang23}, 
7: \citet[][DUALZ]{Fujimoto23b}, 
8: \citet[][UNCOVER, DR4]{Price24},    
9: F. Bauer et al., in prep (ALCS).
(7) Magnification (using the updated v1.1 model 
of \citealt{Furtak23}). 
(8) Stellar mass from Prospector as in \citet{Wang23}, with fixed redshift when $z_{\rm spec}$ or $z_{\rm griz}$ is available and using alternative deblended photometry for noted objects (see Sec~\ref{sec:data}). 
(9) Residual-corrected major axis effective radius measured in F444W, uncorrected for lensing, in arcsec.   
(10) \galfit S\'ersic index in F444W. 
(11) Concentration parameter measured from the F444W residual corrected flux profiles.
(12) Ratio of long to short wavelength residual-corrected 
effective radii. By default, 
F444W and F150W are used for the LW and SW radii, respectively. If F150W has insufficient SNR or a flagged fit, F200W or next F277W is used instead
(noted with \emph{a}).
\emph{Annotation:} 
*: Alternative detection and segmentation used for morphological analysis and stellar population modeling (see Sec.~\ref{sec:data}). 
}
\vspace{-0.85cm}
\end{deluxetable*}


\appendix

\begin{figure*}
\vspace{-1.5mm}
\centering
\includegraphics[width=0.85\textwidth]{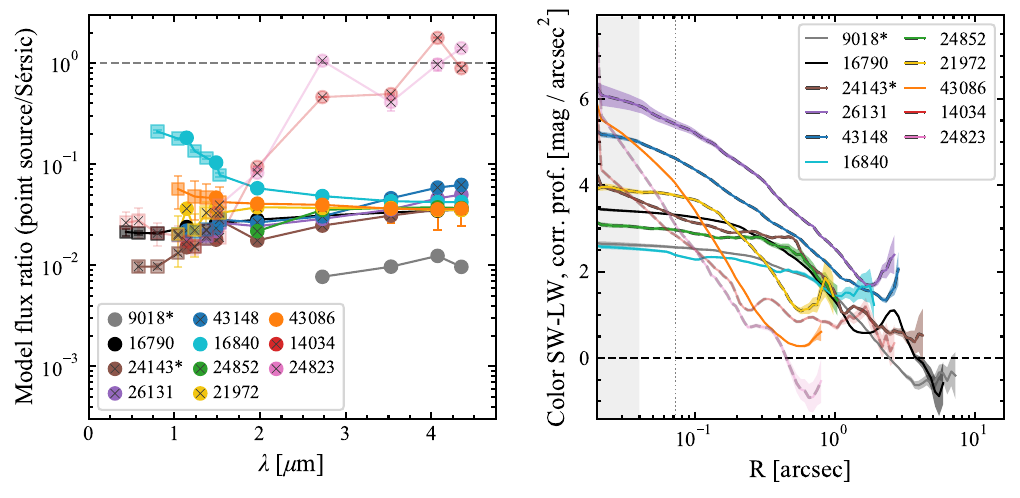}
\vglue -4mm
\caption{
\emph{Left:} 
Ratio of the total integrated fluxes of the point source to the S\'ersic component of the two-component alternative morphological fits, as a function of observed wavelength. 
The marker definitions are the same as in Fig.~\ref{fig:structures_meas}, and X-ray detected objects are marked with gray Xs. 
Amongst the X-ray-detected sources, 
an appreciable ($\gtrsim10\%$) point source flux contribution is only seen for 14034 \& 24823 at long wavelengths ($\gtrsim2\um$), suggesting these 2 objects host dust-reddened but not completely obscured AGN. We infer the host galaxies dominate the morphologies of the other 5 X-ray-detected objects even out to $4.4\um$ (with point sources only $\lesssim6\%$ as bright as the extended component, or a difference of 3 magnitudes and greater). 
\emph{Right:} 
Short-to-long wavelength 
color gradients, derived from 
residual-corrected deconvolved surface-brightness profiles from single S\'ersic component fits (as in Fig.~\ref{fig:SW_LW_fits}). 
The region smaller than the long wavelength pixel scale is shaded gray, and the F444W PSF half-width half-maximum is marked with the vertical dotted line. 
X-ray-detected objects are denoted with an underlying thick gray dashed line. 
All objects have red centers, and continue to exhibit red colors out to large radii. 
The X-ray-detected objects with no appreciable AGN flux contribution even at long wavelengths (\emph{left}; 24143*, 26131, 43148, 24852, 21972) do not exhibit any dramatic change in color profile at the PSF scale, suggesting the AGN is not drastically impacting the overall color gradient or short-to-long wavelength size ratios.}
\vspace{2mm}
\label{fig:appendix}
\end{figure*}

\section{Evaluation of point-source flux contributions from AGN in X-ray-detected objects}
\label{sec:appendixA}

To investigate the impact of AGN on the morphologies of the 7 X-ray detected objects in our sample (24143*, 26131, 43148, 24852, 21972, 14034, 24823), 
we perform an alternative set of morphological fits with \galfit, including a point source as well as a  S\'ersic component. 
These fits are performed using the same methodology as the fiducial, single S\'ersic component fits (as described in Sec.~\ref{sec:methods}). 
The total magnitude of the point source component is free, and its location $x_c,y_c$  is tied or fixed to coincide with the center of the S\'ersic component (depending on whether the S\'ersic center $x_c,y_c$ is free or not for the specific filter, as noted in the procedure).

The measured flux ratio between the point source flux and the total S\'ersic component flux (Fig.~\ref{fig:appendix}a) 
is typically very small, with the point source $\lesssim6\%$ as bright as the S\'ersic component (or $\gtrsim3$ magnitudes fainter). 
Appreciable (assumed to be $\gtrsim10\%$) point source flux contributions 
among the X-ray-detected sources 
are only seen for 14034 \& 24823, and only at long wavelengths ($\gtrsim2\um$). 
(A non-X-ray-detected source, 16840, shows elevated point source flux at short wavelengths $\lesssim1.5\um$, which we attribute to an opportunely located light clump and patchy dust, which fades in prominence towards less-attenuated longer wavelengths.) 
Overall, we find that the S\'ersic parameters derived for these alternative two-component fits are very similar to those from the fiducial, single component fits 
(e.g., median ratio of S\'ersic component $R_{E,{\rm two\, comp}}/R_{E,{\rm one\, comp}}=1.007$), further supporting the lack of impact of a point source to the morphologies for most objects at most wavelengths.

We also examine the residual-corrected, intrinsic short-to-long wavelength color profiles of the sample for indications of AGN-dominated central colors that differ from the overall galaxy. 
We measure residual-corrected surface brightness profiles using the single-component S\'ersic fits (as in Fig.~\ref{fig:SW_LW_fits}), 
measuring the residuals within elliptical annuli 
using the same PA and axis ratio based on the F277W best-fit values for both the SW (F150W, or F200W or next F277W if insufficient SNR in F150W) and the LW (F444W) residual images. 
All objects demonstrate negative color gradients, with very red centers, continuing to less red colors out to the outskirts. 
Overall, we observe no obvious change in color profile at the PSF scale for the 5 X-ray-detected objects without appreciable point source flux contributions (24143*, 26131, 43148, 24852, 21972). 
This suggests that the inferred highly-obscured AGN in these objects 
do not dominate the color gradients, 
or the measured short-to-long wavelength size ratios.

When combined with the color profiles, 
we infer that the host galaxies dominate the light in 5 of the X-ray-detected sources (24143*, 26131, 43148, 24852, 21972), and that these objects likely host highly obscured AGN (given lack of a point source component even at $4.4\um$). 
As a point source contributes $\gtrsim10\%$ of the flux at $\gtrsim2\um$ for 14034 \& 24823, we infer these remaining X-ray-detected objects host a dust-reddened, but not fully obscured AGN. 
As accurately detangling the host galaxy morphology from the AGN contribution can be challenging, we choose to exclude these objects from the detailed morphological analysis in this work. 
We thus also use parameters derived from the fiducial, single-component S\'ersic fits (reflecting the host galaxy of the remaining X-ray-detected objects) for all analysis.

\bibliography{refs}{}
\bibliographystyle{aasjournal}

\end{document}